\documentstyle [12pt]{article}
\newcommand{\be}{\begin{equation}}
\newcommand{\ee}{\end{equation}}
\newcommand{\bea}{\begin{eqnarray}}
\newcommand{\eea}{\end{eqnarray}}
\newcommand{\ba}{\begin{array}}
\newcommand{\ea}{\end{array}}

\begin{document}
\baselineskip = 18 pt
 \thispagestyle{empty}
 \title{
\vspace*{-2.5cm}
\begin{flushright}
\begin{tabular}{c c c c}
\vspace{-0.3cm}
& {\normalsize MPI-Ph/93-103}\\
\vspace{-0.3cm}
& {\normalsize CERN-Th.7163/94}\\
\vspace{-0.3cm}
& {\normalsize February 1994}
\end{tabular}
\end{flushright}
\vspace{1cm}
Electroweak Symmetry Breaking and\\
  Bottom-Top Yukawa
Unification
\\ }
 \author{
 M. Carena$^a$, $\;\;\;$ M. Olechowski$^c$\\
{}~\\
S. Pokorski$^b$\thanks{On leave of absence from the
Institute of Theoretical Physics, Warsaw University}
        $\;$ and  C. E. M. Wagner$^a$\\
 ~\\
{}~\\
$^a$CERN Theory Division,\\
1211 Geneva 23, Switzerland\\
{}~\\
$^b$Max-Planck-Institut
f\"{u}r Physik, Werner-Heisenberg-Institut\\
F\"{o}hringer Ring 6, D-80805 M\"{u}nchen, Germany.\\
{}~\\
$^c$Institute of Theoretical Physics, Warsaw University\\
ul. Hoza 69, 00-681 Warsaw, Poland\\
{}~\\
 }
\date{
\begin{abstract}
The condition of unification of gauge couplings in the minimal
supersymmetric standard model provides successful predictions
for the weak mixing angle as a function of the strong gauge coupling
and the supersymmetric threshold scale. In addition, in
some scenarios, e.g.\ in  the minimal
SO(10) model, the tau lepton and the bottom and top quark
Yukawa couplings unify at the grand unification scale.
The condition of Yukawa unification leads naturally to large
values of $\tan\beta$, implying a
proper top quark--bottom quark mass hierarchy.  In this work,
we investigate the feasibility of unification of the Yukawa couplings,
in the framework
of the minimal  supersymmetric standard model with (assumed)
universal mass parameters at the unification scale and with
radiative breaking of the electroweak symmetry. We show that
strong correlations between the parameters $\mu_0$, $M_{1/2}$ and
$\delta = B_0 - (6r/7) A_0$ appear within this scheme, where
$r$ is the ratio of the top quark Yukawa coupling to its infrared
fixed point value. These correlations have relevant
implications for the sparticle spectrum, which presents
several characteristic features. In addition, we show that
due to large corrections to the running bottom quark
mass induced through the supersymmetry breaking sector of
the theory, the predicted top quark mass and $\tan\beta$ values
are significantly
lower than those previously estimated in the literature.
\end{abstract}}
\maketitle
\newpage
\section{Introduction}
The minimal supersymmetric standard model provides a well
motivated and predictive extension of the successful standard
model of the strong and electroweak interactions.
The condition
of unification of couplings is implicit within this scheme and
the predictions for the weak mixing angle are in  good
agreement with the values measured by the most recent
measurements at LEP \cite{unif}--\cite{LP}.
In addition to the gauge coupling unification
condition, relations between the values of the Yukawa couplings
of the quarks and leptons of the third generation appear in the
minimal supersymmetric grand unification scheme. In particular,
in the minimal SU(5) model the unification of bottom and tau
Yukawa couplings is obtained. The bottom--tau Yukawa unification
condition leads to predictions for the top quark mass as a
function of the running bottom quark mass, the strong gauge
coupling value and the value of $\tan\beta$, the ratio of
vacuum expectation values \cite{Ramond}--\cite{DHR}.

Recently, it has been observed that for the phenomenologically
allowed values of the bottom quark mass and moderate values
of $\tan\beta < 10$, large values of the top
quark Yukawa coupling are needed in order to contravene the
strong gauge coupling renormalization of the bottom Yukawa
coupling \cite{BABE}--\cite{LP2}.
In general, for large enough values of the top quark Yukawa
coupling at the grand unification scale, the low energy Yukawa
coupling is strongly focussed to a quasi infrared fixed
point \cite{IR}--\cite{Dyn}.
In the minimal supersymmetric standard model, the quasi infared
fixed point predictions for the physical top quark mass $M_t$
are given by $M_t \simeq A \sin\beta$, with $A \simeq 190 - 210$
GeV for the strong gauge coupling $\alpha_3(M_Z) = 0.11 - 0.13$.
It has been recently shown that for the values of
the strong gauge coupling consistent with the
condition of gauge
coupling unification,
with reasonable threshold
corrections at the grand unification and supersymmetry breaking
scales, the top quark mass should be within $10\%$ of its
quasi infrared fixed point values
if the condition of
bottom--tau Yukawa unification is required \cite{BCPW}.

A more predictive scheme is obtained in the framework of the
minimal SO(10) unification.
In this case top--bottom quark
Yukawa unification is also required, implying that,
for a given value of the bottom quark mass and the
strong gauge coupling value \cite{ALS},
not only
the top quark mass but also the value of $\tan\beta$ may be
determined. Remarkably,  large values of
$\tan\beta \geq 40$ are obtained in this case, leading to a
proper bottom--top mass hierarchy
\cite{Banks}--\cite{Hall1}.
For these large values
of $\tan\beta$, the bottom quark Yukawa coupling itself
plays a relevant role in the running of the top quark Yukawa
coupling, as well as
in the running of the ratio of the bottom to
tau Yukawa couplings. This leads to a somewhat weaker
convergence of the top quark Yukawa coupling to its infrared
fixed point value, together with a slight modification of
the infrared fixed point expression \cite{CPW}.

Moreover, it has been recently  observed that for these
large values of $\tan\beta$, potentially large corrections to
the running bottom quark mass may be induced through
the supersymmetry breaking sector of the theory \cite{Hall} --
\cite{Ralf}.
Although in the exact supersymmetric theory the bottom
quark and tau lepton
only couple to one of the Higgs fields $H_1$, a
coupling of these fermions
to the Higgs field $H_2$ is induced at the one loop level
in the presence of soft supersymmetry breaking terms.
These corrections are
 decisive in obtaining  the predictions for the
top quark mass.   Indeed, for
the characteristic values of $\tan\beta$ arising
if the top--bottom Yukawa unification is required,
$\tan\beta = {\cal{O}}(50)$, the bottom mass
corrections  would be very large,
unless the supersymmetric mass parameter $\mu$ and
the gluino mass are much lower than the characteristic
squark masses. This hierarchy of masses
may be achieved by imposing certain symmetries in the theory.
These symmetries may, however,
be in conflict with the radiative breaking of the
electroweak symmetry, particularly in simple supersymmetry
breaking scenarios.

The question of radiative electroweak symmetry breaking with
large $\tan\beta$ appears hence, as an independent issue, which
has been investigated in  minimal supersymmetric models
with universal soft supersymmetry breaking parameters at the
grand unification scale with encouraging results
\cite{OP2}, \cite{SO10}, \cite{AS}, \cite{OP}.
However,
not enough attention was paid either to the full  consistency
with the requirement of the unification of the gauge and
Yukawa couplings, nor to a
systematical  identification of the complete
parameter space at the GUT scale, which gives electroweak symmetry
breaking with large $\tan\beta$. Recently, we presented
an investigation of the properties of the radiative electroweak
symmetry breaking solutions for small and moderate values of
$\tan\beta$ and a top quark Yukawa coupling taking
values close to its infrared fixed point solution, as required
by bottom--tau Yukawa coupling unification \cite{COPW}.
We obtained quite
remarkable correlations between different supersymmetric mass
parameters, as well as an effective reduction of the number
of free independent parameters at the grand unification scale.
It is the purpose of this work
to perform a similar analysis for the large $\tan\beta$ regime.

We use the recently developed bottom -
up approach to radiative electroweak symmetry breaking, \cite{OP},
which is
particularly suitable for a systematic search for large $\tan\beta$
solutions, and possibly to identify the symmetries underlying
those solutions.
In our calculation we use
the two loop renormalization
group evolution of gauge and Yukawa couplings, while the Higgs
and supersymmetric mass parameters are evolved at the one loop
level.
The leading supersymmetric threshold corrections to the Higgs quartic
couplings
and to all supersymmetric mass parameters
are included in the analysis. We proceed by fixing
the experimentally known values of $M_Z$, $\alpha(M_Z)$,
$\sin^2\theta_W(M_Z)$ and $M_{\tau}$ (with their
corresponding uncertainties).
After choosing a set of values for
$M_t$ and $\tan\beta$,
the unification condition of the three
Yukawa couplings
fixes their running in the range from $M_Z$ to $M_{GUT}$.
Next,
the search for electroweak symmetry
breaking solutions is performed by
scanning over the CP odd Higgs mass and the low energy
stop mass parameters.
For each solution the one--loop correction to the running
bottom mass at $M_Z$ is calculated and finally the pole
bottom mass is obtained. The predictions for the top quark mass
and $\tan\beta$ is the collection of those values of
$M_t$ and $\tan\beta$ for which there are solutions with
the pole bottom mass within the experimentally acceptable range.
A more detailed explanation of this procedure
will be given below.

We will show that
under the requirement of  top--bottom--tau Yukawa unification,
the condition of radiative electroweak symmetry
breaking implies strong correlations
between the supersymmetric
parameter $\mu_0$ and the soft supersymmetry breaking term
$M_{1/2}$ and $\delta = B_0 - (6r/7) A_0$, where $r$ is the
ratio at the electroweak scale
of the top quark Yukawa coupling to its infrared fixed
point value. These correlations allow a precise determination
of the bottom mass corrections, which become significantly
large for the large values of $\tan\beta$ consistent with
the unification of the three Yukawa couplings of the third
generation. This, in turn,
implies that the top quark mass predictions are quite
different
from those ones obtained if the bottom mass corrections were
neglected.
In section 2 we present a general discussion of the model and
our choice of low energy parameters.
In section 3 we discuss the radiative electroweak symmetry
breaking conditions and its implication for the low energy
parameters of the Higgs potential. We present an approximate
analytical expression for the one loop renormalization group
running of the mass parameters of the Higgs and supersymmetric
particles, for the case in which the top and bottom Yukawa
couplings unify at $M_{GUT}$. In section 4 we present a
detailed numerical  analysis of the implications
of the radiative $SU(2)_L \times U(1)_Y$
breaking for the
supersymmetric mass parameter $\mu$ and the supersymmetry
breaking parameters at the grand unification scale, and
we compare it with the approximate analytical solution.
In section 5 we analyse the one loop corrections to the bottom
and tau masses and their implications
for the top quark mass
predictions. In section 6 we analyse the spectrum, together
with the constraints coming from the bounds on the
$b \rightarrow s \gamma$ decay rate. We reserve section 7 for the
conclusions.

\section{Gauge and Yukawa Coupling Unification Predictions}

We begin with a short discussion of the predictions for the
top quark mass following from the unification of the gauge
and Yukawa couplings
(before imposing the requirement of radiative electroweak
breaking),
recalling and slightly extending some of
the results presented
in Refs. \cite{LP} - \cite{BCPW}.
The gauge coupling unification condition gives predictions
for the weak mixing angle $\sin^2\theta_W(M_Z)$ as a function
of the strong gauge coupling $\alpha_3(M_Z)$. The unification
condition implies (at the two loop level) the following
numerical correlation \cite{BCPW}
\begin{equation}
\sin^2\theta_W(M_Z) = 0.2324 - 0.25 \left(
\alpha_3(M_Z) - 0.123 \right) \pm 0.0025
\end{equation}
where the central value corresponds to an effective supersymmetric
threshold scale $T_{SUSY} = M_Z$ and the error $\pm 0.0025$
is the estimated uncertainty in the prediction arising from
possible supersymmetric threshold corrections (corresponding to
vary the effective supersymmetric threshold scale $T_{SUSY}$ from
15 GeV to 1 TeV), threshold corrections at the unification
scale as well as from higher dimensional operators. On the
other hand, $\sin^2\theta_W(M_Z)$ is given by the
electroweak parameters $G_F$, $M_Z$, $\alpha_{em}$ as a
function of the physical top quark mass
$M_t$ (at the one loop level) by the formula \cite{LP}:
\begin{equation}
\sin^2\theta_W(M_Z) = 0.2324 - 10^{-7} GeV^{-2}
\left( M_t^2 - (138 GeV)^2 \right) \pm 0.003
\end{equation}
Therefore, the predictions from gauge coupling unification agree
with experimental data provided
\begin{equation}
M_t^2 = (138 GeV)^2 +  0.25 \times 10^7 GeV^2
\left( \alpha_3(M_Z) - 0.123 \pm 0.01 \right)
\label{eq:Mtunif}
\end{equation}
The above $M_t - \alpha_3(M_Z)$ correlation defines a band
whose lower bound is shown in Fig. 1 ( the upper bound is
above 0.13 for $M_t > 110$  GeV). We observe that the top quark
mass $M_t > 110 \; (155) GeV$
implies $\alpha_3(M_Z) > 0.11 \; (0.115)$.

Another issue is that of unification of the bottom and tau
Yukawa couplings.
In this work  the unification of
Yukawa couplings  is always studied numerically
at the two--loop level. However, for a qualitative
discussion we refer to
the one--loop renormalization group equation for the ratio
of the bottom to tau Yukawa couplings $r = h_b/h_{\tau}$
which, in
the limit of vanishing electroweak gauge couplings, reads
\begin{equation}
\frac{dr}{dt} = \frac{r}{8\pi}
\left( \frac{16 \alpha_3}{3} - 3 Y_b - Y_{t} + 3 Y_{\tau} \right)
\label{eq:r}
\end{equation}
where $t = \ln(M_Z/Q)^2$ and $Y_i = h_i^2/(4\pi)$.
Starting from values of the ratio
$r$ above one at the scale $M_b$, as required by experimentally
allowed values of the bottom mass
$M_b = (4.9 \pm 0.3)$ GeV \cite{Pdat}  and
the tau mass, $M_{\tau} = 1.78$ GeV, $r$ is strongly renormalized
and in the limit of negligible Yukawa couplings for values of
$\alpha_3$ within the experimentally determined range it
becomes lower than one at scales far below the grand unification
scale $M_{GUT}$. Hence, in order to get $r(M_{GUT}) = 1$, for
a given value of $\alpha_3$, the Yukawa couplings in
Eq.(\ref{eq:r}) should be adjusted to compensate the strong gauge
coupling effect. For low and
moderate values of $\tan\beta$,
$\left( Y_b, Y_{\tau}  \ll Y_{t} \right)$, it is
the top quark Yukawa coupling which is fixed as a function
of $\alpha_3$, by the bottom--tau Yukawa
unification requirement. As we discussed
in the introduction, this perturbative unification requires
values that are within  10$\%$ of the top quark mass infared
quasi fixed point value.

Here we are primary concerned with
the large $\tan\beta$ solution.
Then, the bottom and the top quark Yukawa couplings are of the
same order of magnitude and both are important to get
$r(M_{GUT}) \simeq 1$. The unification of the three Yukawa couplings
takes place not only for a particular value of $Y_t$ but also
of $\tan\beta$, for given values of $M_b$, $M_{\tau}$
and $\alpha_3(M_Z)$, implying a fixed $M_t$.
An important
remark is in order here. The bottom mass which is directly relevant
for the top mass prediction following from the Yukawa coupling
unification is the tree level running mass $m_b(M_z)$.
As we discussed in the previous
section, in the large $\tan\beta$ case it
may receive large loop corrections from sparticle
exchange loops, at least in some range of parameters of the model.
The physical (pole mass) $M_b$ is obtained from the running mass
$m_b(M_b)$ (which is related to $m_b(M_Z)$ by the Standard Model
RG equations) by inclusion of
QCD corrections, which are universal for the Standard Model
and its supersymmetric version.
At the two loop level,
they are given by \cite{Ramond2}
\begin{equation}
m_b(M_b) = \frac{M_b}{1 + \frac{4\alpha_3(M_b)}{3 \pi} +
K_b \left( \frac{\alpha_3(M_b)}{\pi} \right)^2 },
\label{eq:mb}
\end{equation}
where $K_b = 12.4$.
The loop corrections to the running bottom mass at $M_Z$
induced through
sparticle exchange loops are an important issue
for models with radiative
breaking of the electroweak symmetry.
In order to distinguish them
from QCD corrections, we introduce the pole bottom mass
$\tilde{M}_b$, which
is obtained from the unification condition
in the case in which the supersymmetric one loop corrections
to $m_b(M_Z)$ are ignored. Due to the fact that the
supersymmetric corrections could be quite sizeable,
for the allowed solutions of the model,
the mass $\tilde{M}_b$ may be significantly
different from the physical mass $M_b$.

The predictions for $M_t$ and $\tan\beta$, following from the
unification of the three Yukawa couplings, are
shown in Fig. 1 for several values of
the mass
$\tilde{M}_b$
as a function of $\alpha_3(M_Z)$.
The supersymmetric particle
masses were set at the scale $M_Z$, while the unification scale
was defined as the scale at which the electroweak gauge couplings
unify. Fig. 1 shows also the region in the $\alpha_3(M_Z)$ -
$M_t$ plane consistent with the unification of gauge
couplings, after
considering the experimental dependence of $\sin^2\theta_W(M_Z)$
on the top quark mass and threshold corrections at the
supersymmetric and grand unification scale, Eq.(\ref{eq:Mtunif}).

{}From Fig. 1  we draw the following conclusions:
In  case that the supersymmetric loop contributions
to the bottom mass were negligible, $\tilde{M}_b = M_b$, and
taking into account the experimentally acceptable values for
the physical bottom mass,
$M_b \simeq 4.9 \pm 0.3$ GeV \cite{Pdat}, the
unification of the gauge and Yukawa couplings drives the top
quark mass towards large values (Note the fact that for
$Y_b = Y_{t}$ the IR fixed point solution is lower than for
$Y_b \ll Y_t$) \cite{CPW}.
Although the predictions for $M_t$ are
no longer so strongly constrained to be close to its
infrared quasi fixed point values as for the low and moderate
values of $\tan\beta$ (as explained above, strong
renormalization effects in the running of $h_b$ are
partially cancelled by $h_b$ itself), for the values of
$\alpha_3(M_Z)$ consistent with gauge coupling unification
the top quark mass is still close  to the appropriate infrared
fixed point solution. For instance, for a
physical bottom quark mass $M_b = 5.2$ GeV,
$\alpha_3(M_Z) \simeq 0.12$ and $\tan\beta = 50$, the top
quark mass is predicted to be $M_t \simeq 175$ GeV.
In general,
as it is clear from Fig. 1,
if the supersymmetric corrections to the running bottom mass
were small, the top quark mass would
acquire values $M_t > 165$ GeV
within this scheme \cite{Hall}. In Fig. 1 we also plot the
predictions obtained for
values of $\tilde{M}_b$ larger than the experimental upper bound
for the bottom mass, $M_b < 5.2$ GeV, which
will become of interest while studying the supersymmetric
corrections to the bottom mass.  Indeed,
the values of the top quark mass in the range,
say (140--160) GeV are compatible with unification of couplings
provided $\tilde{M}_b > M_b$ and sizeable supersymmetric
loop corrections to the bottom mass are induced.
As we shall show
below, this is the case in the minimal supergravity model with
minimal $SO(10)$ Yukawa unification.
It is relevant to contrast this situation with what happens
for low and moderate values of $\tan\beta$, for which the
consistency of a moderately heavy top quark, $M_t < 160$ GeV,
with bottom--tau Yukawa
unification requires the ratio of vacuum expectation
values to be very close to one, $\tan\beta < 1.4$, unless
large threshold corrections to both gauge and Yukawa couplings
are present at the grand unification scale \cite{BABE},
\cite{LP2},\cite{BCPW}.

\section{Higgs Potential Parameters}

In order to analyze the radiative electroweak symmetry
breaking condition, one should concentrate on the Higgs
potential of the theory. In the
Minimal Supersymmetric
Standard Model, and after the inclusion of the
leading--logarithmic
radiative corrections, it
may be written as \cite{Dyn}, \cite{CSW}--\cite{HH}
\begin{eqnarray}
V_{eff} & = & m_1^2 H_1^{\dagger} H_1 +
m_2^2 H_2^{\dagger} H_2 - m_3^2 (H_1^T i \tau_2 H_2
+ h.c.)
\nonumber\\
& + & \frac{\lambda_1}{2} \left(H_1^{\dagger} H_1 \right)^2
+ \frac{\lambda_2}{2} \left(H_2^{\dagger} H_2 \right)^2
+ \lambda_3 \left(H_1^{\dagger} H_1 \right)
 \left(H_2^{\dagger} H_2 \right)
+ \lambda_4 \left| H_2^{\dagger} i \tau_2 H_1^* \right|^2
\end{eqnarray}
where the quartic couplings may be obtained by the
corresponding renormalization group equations and
the fact that,
at scales at which the theory is supersymmetric  the
running quartic couplings $\lambda_j$, with $j = 1 - 4$,
must satisfy the following conditions
Refs. \cite{CSW}--\cite{Chankowski}:
\begin{equation}
\lambda_1 = \lambda_2 = \frac{ g_1^2 + g_2^2}{4},\;\;\;\;\;
\lambda_3 = \frac{g_2^2 - g_1^2}{4},\;\;\;\;\;
\lambda_4 = - \frac{g_2^2}{2}.
\end{equation}
  The masses $m_i^2$, with $i = 1-3$ are also running mass
parameters, whose renormalization group equations may be
found in the literature \cite{Savoy}-\cite{Ibanez}. As we
explained in section 1, in the numerical analysis we considered
the two loop renormalization group evolution of gauge and
Yukawa couplings, while the supersymmetric and Higgs mass
parameters, as well as the low energy Higgs quartic couplings
are evolved at the one loop level
with the leading supersymmetric threshold corrections included.
The minimization conditions read
\begin{equation}
\sin(2\beta) = \frac{ 2  m_3^2  }{m_A^2} ,
\label{eq:s2b}
\end{equation}
\begin{equation}
\tan^2\beta = \frac{m_1^2 + \lambda_2 v^2 +
\left(\lambda_1
 - \lambda_2 \right) v_1^2}{m_2^2 + \lambda_2 v^2} ,
\label{eq:tb}
\end{equation}
where  $\tan\beta = v_2/v_1$, $v_i$ is  the vacuum expectation
value of the Higgs fields $H_i$, $v^2 = v_1^2 + v_2^2$,
and $m_A$ is the CP-odd Higgs
mass,
\begin{equation}
m_A^2 = m_1^2 + m_2^2 + \lambda_1 v_1^2 +
\lambda_2 v_2^2 + \left( \lambda_3 + \lambda_4 \right) v^2
\end{equation}
and we define  the mass parameter $m_3^2$ to be positive.

Apart from the mass parameters $m_i^2$, appearing in the
effective potential, the evolution of the supersymmetric
mass parameter $\mu$ appearing in the superpotential $f$,
\begin{equation}
f =  h_t \epsilon_{ij} Q^j U H_2^i
+ h_b \epsilon_{ij} Q^i D H_1^j
+ h_\tau \epsilon_{ij} L^i E H_1^j
+ \mu \epsilon_{ij}
H_1^i H_2^j ,
\end{equation}
(where $Q^\top = (T\;B)$ is the top--bottom left handed
doublet superfield
and $U$, $D$ and $L$ are $SU(2)_L$
singlet superfields) is relevant
for the analysis of the radiative
electroweak symmetry breaking conditions.  The bilinear mass
term proportional to $m_3^2$ appearing in the Higgs potential may
be rewritten as a soft supersymmetry breaking parameter $B$ multiplied
by the Higgs bilinear term appearing in the superpotential, that
is $m_3^2 = B \mu$. Analogously, the scalar potential may contain a
scalar trilinear breaking term  proportional to the $h_f$ -
Yukawa dependent
part of the superpotential, with a trilinear coupling $A_f$.

In order to get an understanding of the numerical
results, we will present approximate analytical formulae,
for the relations required by
the electroweak symmetry breaking
conditions, in
which the radiative corrections to the quartic couplings
are ignored.
There are several features of the Higgs potential which are
characteristic for large $\tan\beta$ values. They can be
easily discussed in a qualitative way on the basis of
the supersymmetric tree level potential.
Eq.(\ref{eq:tb}) simplifies to
\begin{equation}
\tan^2\beta = \frac{ m_1^2 + M_Z^2/2}{ m_2^2 + M_Z^2/2} \; ,
\label{eq:tb2}
\end{equation}
so, for large $\tan\beta$ (already, say, $\tan\beta > 30$),
either
\begin{equation}
m_2^2 \simeq - \frac{M_Z^2}{2} \;,
\label{eq:m1m21}
\end{equation}
if $m_1^2, m_2^2$ are of the order of the $Z^0$ boson mass
squared, or
\begin{equation}
m_1^2 \simeq \tan^2\beta m_2^2
\label{eq:m1m2}
\end{equation}
when $m_1^2,m_2^2 \gg M_Z^2$. In general, the smaller is the
cancellation in the denominator of Eq.(\ref{eq:tb2}), the larger is the
hierarchy between $m_1^2$ and $m_2^2$.
The second relation, Eq.(\ref{eq:m1m2}), is, however,
unnatural
when $Y_b \simeq Y_t$. Indeed, if all supersymmetric particle masses
are below a few TeV, Eq.(\ref{eq:m1m21}) holds, within a good
approximation (Although the inclusion of radiative corrections
modifies the low energy convergence of the $m_2^2$ parameter,
the relation $|m_2^2| \approx\frac{1}{2} M_Z^2$ is preserved, what is
sufficient for the understanding of the properties discussed
below). Eq.(\ref{eq:m1m21}), combined with the
condition $M_A^2 \simeq m_1^2 + m_2^2 > 0$, gives a useful
constraint:
\begin{equation}
m_1^2 - m_2^2 > M_Z^2.
\label{eq:diff1}
\end{equation}
Another very important property is
\begin{equation}
m_3^2 \simeq \frac{M_A^2}{\tan\beta},
\label{eq:m32}
\end{equation}
or, equivalently, $m_1^2 \gg m_3^2$. Since in the $Y_t \simeq
Y_b$ case a large hierarchy between $m_1^2$ and $m_2^2$ is
highly unnatural,
 the above condition, Eq.(\ref{eq:m32}), implies
also $|m^2_2|\gg m^2_3$ ($M_Z^2 \gg m^2_3$).
Thus, in order to study the implication of the electroweak
symmetry breaking condition, one can effectively replace
Eq.(\ref{eq:m32}) with the condition  $m_3^2 \simeq 0$.

To go further with the analysis, it is very useful to obtain
approximate analytical solutions for the one loop renormalization
group evolution of the mass parameters, whose validity may be
proven by comparing them with our numerical solutions.
We will assume
universality of the soft supersymmetry breaking parameters,
that is to say a common scalar mass $m_0$ and a common gaugino
mass $M_{1/2}$, as well as the
boundary conditions  for the parameters
$A_t$ ($A_b$ and $A_{\tau}$),
$B$ and $\mu$, at the grand unification scale to be given by
$A_0$, $B_0$ and $\mu_0$, respectively.
In the region of large values  of $\tan\beta$, for which
the bottom  Yukawa coupling is of the order of the top Yukawa
coupling, an approximate analytical solution for the one loop
evolution of the mass parameters may be obtained. For this, we
identify the bottom and top Yukawa couplings
and neglect the tau Yukawa coupling effects.
Furthermore, all supersymmetric threshold corrections
are ignored at this level. The solution
for $Y = Y_{t} \simeq Y_b$  reads
\begin{equation}
Y_t(t) = \frac{ 4 \pi Y_t(0) E(t)}{ 4 \pi + 7 Y_t(0) F(t)}
\end{equation}
where $E$ and $F$ are functions of the gauge couplings,
\begin{equation}
E = (1 + \beta_3 t)^{16/3b_3}
(1 + \beta_2 t)^{3/3b_2}
(1 + \beta_1 t)^{13/9b_1},
\;\;\;\;\;\;\;\;\;\;\;\; F= \int_{0}^t E(t') dt'
\end{equation}
 with $\beta_i = \alpha_i(0) b_i/4\pi$, $b_i$     the
beta function coefficient of the gauge coupling $\alpha_i$,
$t = 2 \log(M_{GUT}/Q)$ and we identify the right bottom and the
right top hypercharges. As we
said, the fixed point solution is obtained for values of the
top quark Yukawa coupling which become large at the grand
unification scale, that is, approximately,
\begin{equation}
Y_f(t) = \frac{4\pi E(t)}{7 F(t)}.
\label{eq:fixi}
\end{equation}
As had been anticipated
in Ref.\ \cite{CPW}, the
fixed point solution for the $Y_b \simeq Y_t$ case differs
in a factor $6/7$ from the corresponding solution in the
low $\tan\beta$ case, for which $Y_t \gg Y_b$.
{}From here, by inspecting the renormalization group
equation for the mass parameters, we obtain the
approximate analytical solutions
\begin{equation}
m_{H_1}^2 \simeq m_{H_2}^2 = m_0^2 + 0.5 M_{1/2}^2 -
\frac{3}{7} \Delta m^2
\label{eq:m12}
\end{equation}
\begin{equation}
m_U^2 \simeq m_D^2 = m_0^2 + 6.7 M_{1/2}^2 - \frac{2}{7}
\Delta m^2
\label{eq:sqm2}
\end{equation}
\begin{equation}
m_Q^2 \simeq  m_0^2 + 7.2 M_{1/2}^2 - \frac{2}{7}
\Delta m^2
\label{eq:sqm}
\end{equation}
where  $m_i^2 = \mu^2 + m_{H_i}^2$, with $i = 1,2$, $m_Q^2$,
$m_D^2$, $m_U^2$ are the squark doublet, right bottom squark
and right stop quark mass parameters respectively and
\begin{eqnarray}
\Delta m^2 &  \simeq
&  3 m_0^2 \frac{Y}{Y_f} - 4.6 A_0 M_{1/2}
\frac{Y}{Y_f} \left( 1 - \frac{Y}{Y_f} \right)
\nonumber\\
& + &
A_0^2 \frac{Y}{Y_f} \left( 1 - \frac{Y}{Y_f} \right)
+ M_{1/2}^2
\left[
14 \frac{Y}{Y_f} - 6
\left(
\frac{Y}{Y_f} \right)^2 \right] \; .
\label{eq:dm}
\end{eqnarray}
Here we have concentrated on the above mass parameters, because
they are the only relevant ones
for the study of the properties of
the radiative electroweak symmetry breaking solutions
in the approach of ref.~\cite{OP}. We
will discuss the properties of the mass spectrum in more
detail in section 6.
Moreover, the supersymmetric mass parameter renormalization group
evolution gives,
\begin{equation}
\mu^2 = 2 \mu_0^2 \left( 1 - \frac{Y}{Y_f} \right)^{6/7}  \; ,
\label{eq:mu}
\end{equation}
while the running of the soft supersymmetry breaking bilinear
and trilinear coupling read,
\begin{equation}
A_t = A_0 \left( 1 - \frac{Y}{Y_f} \right) - M_{\frac{1}{2}} \left(
4.2 - 2.1 \frac{Y}{Y_f} \right),
\label{eq:A0}
\end{equation}
\begin{equation}
B \simeq \delta(Y) + M_{1/2} \left( 2
\frac{Y}{Y_f} - 0.6 \right),
\label{eq:b0}
\end{equation}
with
\begin{equation}
\delta(Y) = B_0 - \frac{6 Y}{7 Y_f} A_0.
\end{equation}
 The coefficients  characterizing the
 dependence of the mass parameters on the universal gaugino
mass $M_{1/2}$
are functions of the exact value of the gauge
couplings. In the above, we have taken the values of the coefficients
that are obtained for  $\alpha_3(M_Z)\simeq 0.12$.

The approximate solutions, Eqs. (\ref{eq:m12}--\ref{eq:A0}),
become weakly dependent on the parameter $A_0$, the dependence
being weaker for top quark Yukawa couplings closer to the fixed
point value. The strongest dependence on the parameter $A_0$
comes through the parameter $\delta(Y)$ introduced above.
Similar properties are obtained in
the low  $\tan\beta$ regime \cite{COPW}, although
the explicit form of the parameter $\delta$ is different
in this case. From Eq.(\ref{eq:mu}), it follows that
the coefficient relating $\mu$ to $\mu_0$ tends  to
zero as $Y  \rightarrow Y_f$. The coefficients scales
faster to zero than in the low $\tan\beta$ case.

\section{Radiative Breaking of $SU(2)_L \times U(1)_Y$}

In the following we present a complete numerical analysis of the
constraints coming from the requirement of a proper radiative
electroweak symmetry breaking in the large $\tan \beta$ regime.
As described in the Introduction, we use the bottom--up approach
of ref.~\cite{OP}. For a fixed value of the top quark mass $M_t$
we search for all solutions to radiative breaking, which give
a chosen value of $\tan\beta$, by scanning over the CP odd Higgs
mass and the low energy stop mass parameters. The latter are very
convenient as the input parameters as they fix the leading
supersymmetric threshold corrections to the Higgs potential.
While studying the model from low energies we have chosen for
definiteness an upper bound of 2 TeV on the scanned parameters.
For a somewhat larger upper bound, larger values of the soft
supersymmetry breaking prameters are allowed, but the general
features of the solutions are preserved. It is natural to
expect that the supersymmetric parameters are at most of order
of a few TeV, if supersymmetry is to solve the hierarchy problem
of the Standard Model.
In Figs. 2 - 5
we present the results which show interesting correlations among the
soft supersymmetry breaking parameters.

As discussed in section 3, the one loop corrections to
the effective  Higgs potential, necessary to perform
a proper analysis of the radiative electroweak symmetry
breakdown, were included in the numerical analysis.
The gauge and Yukawa couplings were evolved with
their two loop renormalization group equations between
$M_Z$ and $M_{GUT}$. In their evolution, we have treated
all supersymmetric particle masses as being equal to
$M_Z$. Although this procedure introduces small uncertainties on
the predicted values of $\alpha_3(M_Z)$ and $M_t$ (which
will be considered in our analysis), it keeps all the
essential features of the radiative electroweak symmetry
with unification of bottom and top Yukawa couplings,
makes possible the comparison of our results with the ones
of Fig. 1 and allows an easy analytical interpretation of
the numerical results. In addition, the small
uncertainties on $\alpha_3(M_Z)$ and $M_t$ may be treated  by
analytical methods \cite{CPW}, \cite{Hall1}.

Analogously to the low
$\tan \beta$ scenario \cite{COPW}, it is possible to derive
approximate analytical relations,
which are useful in the understanding of the numerical results.
Indeed, considering the conditions for a proper radiative electroweak
symmetry breaking, Eq.(\ref{eq:m1m21}),
the approximate solutions for the mass parameters,
Eqs. (\ref{eq:m12}) -(\ref{eq:dm}),
and ignoring radiative corrections to the quartic couplings
the following analitycal expression is obtained,
\begin{eqnarray}
\mu^2 & = &  m_0^2 \left( \frac{9}{7} \frac{Y}{Y_f} - 1 \right)
- M_{1/2}^2 \left[ 0.5 - 6 \frac{Y}{Y_f} + \frac{18}{7}
\left(\frac{Y}{Y_f} \right)^2 \right]
\nonumber\\
& - &
 \frac{3}{7} 4.6 A_0 M_{1/2} \frac{Y}{Y_f} \left( 1 - \frac{Y}{Y_f}
\right) +  \frac{3}{7}  A_0^2 \frac{Y}{Y_f} \left( 1 - \frac{Y}{Y_f}
\right)
- M_Z^2/2 \;.
\label{eq:mu2}
\end{eqnarray}
In the analytical presentation
we will always keep the expressions as a
function of the low energy parameter $\mu$. The reason is
that in the one loop approximation $\mu$ and $\mu_0$ are
linearly related, Eq.(\ref{eq:mu}), and $\mu$ becomes
a more appropriate parameter for the description of the
solution properties, particularly for large values of the
top quark mass where $\mu_0$ strongly depends on the degree
of proximity to the fixed point value. The $\mu_0$ dependence
may be always recovered by using Eq.(\ref{eq:mu}).

In the above we have taken the expression of $m_2^2$ obtained
in the analytical approximation in which $m_1^2 \simeq m_2^2$.
In the
 explicit numerical solution to the mass parameters, however,
 we obtain
\begin{equation}
m_1^2 - m_2^2 = \alpha M_{1/2}^2 + \beta m_0^2
\label{eq:diff2}
\end{equation}
where for $Y/Y_f \simeq 1$ ($M_t \simeq 190$ GeV),
$\alpha \simeq 0.2$, and  $\beta \simeq -0.2$, while for
$Y/Y_f \simeq 0.6$ ($M_t \simeq 150$ GeV),
$\alpha \simeq 0.1$ and $\beta \simeq
- 0.1$. Hence, the coefficient $\alpha$ is small and positive,
and $\beta$ is negative and small in magnitude.
The order of magnitude of the coefficients $\alpha$
and $\beta$ can be easily inferred from the renormalization
group equations. Indeed, it is easy to show that
under the condition of unification of the three third generation
Yukawa couplings $\alpha$ comes mainly from the difference
in the running of bottom and top Yukawa couplings, together
with the different hypercharges of the right top and bottom
quarks, which induce a different gaugino dependence of the
stop and sbottom parameters. The negative values of
$\beta$ are mainly due to the $\tau$ lepton Yukawa effects.
We see that,
due to the restriction $m_1^2 - m_2^2 > M_Z^2$, Eq.(\ref{eq:diff1}),
values of $m_0^2 > M_{1/2}^2$ make the radiative breaking of
the electroweak symmetry
impossible, in the approximation which neglects
supersymmetric threshold corrections to the Higgs potential.
In the numerical analysis,
which includes those corrections,
the only solutions
are still obtained for $M_{1/2}^2$ of the order of, or larger than
$m_0^2$, as seen in Fig. 2.a and Fig. 2.b.
It is also important to remark that, due to the
smallness of the parameters $\alpha$ and $\beta$,
the dependence of the mass parameters $m_1^2$ and $m_2^2$ on
the gaugino mass $M_{1/2}$ is well described by Eq.(\ref{eq:mu2})
(which was obtained in the approximation $m^2_1 = m^2_2$),
while the dependence on the mass parameters $A_0$, $m_0$,
remains weak for $M_{1/2} > m_0$.
In general, the corrections to the approximate
solutions given in
Eqs.(\ref{eq:fixi} - \ref{eq:sqm}) and
Eq.(\ref{eq:mu2}) are small, and, hence
they provide useful information for the analysis of the
electroweak symmetry breaking condition.

The values of the top quark mass,
$M_t = 190$ GeV and $M_t = 150$ GeV,
and of the ratio of the Higgs vacuum expectation values,
$\tan\beta = 55$ and $\tan\beta = 38$,
used above are such
that unification of gauge and Yukawa couplings is achieved
for $\tilde{M}_b \simeq 5.4$ GeV,
$\alpha_3(M_Z) \simeq
0.129$ and $\tilde{M}_b \simeq 5.85$ GeV,
$\alpha_3(M_Z) \simeq 0.124$, respectively.
Considering a Yukawa coupling solution sufficiently
close to the infrared fixed point, the values of $M_{1/2}^2
\geq m_0^2$ as required by the radiative breaking conditions,
and $M^2_{1/2}>M^2_Z$ (as follows from
Eqs.(\ref{eq:diff1}, \ref{eq:diff2}),
we obtain from Eq.(\ref{eq:mu2}) that,
\begin{equation}
\mu^2 \simeq
 3  M_{1/2}^2 ,
\label{eq:mum12}
\end{equation}
i.e.\ there is a strong linear correlation between $\mu$ and $M_{1/2}$.
If, instead, we consider the case $Y/Y_f = 0.6$, (corresponding
to $M_t \simeq 150$ GeV) as a representative
one of what happens when we depart from the fixed point value
we obtain
\begin{eqnarray}
\mu^2 & = & -0.23 m_0^2 +2.2 M_{1/2}^2 - 0.47 A_0 M_{1/2}
\nonumber\\
& + & 0.1 A_0^2 - M_Z^2/2.
\end{eqnarray}
There is a stronger dependence on the supersymmetry
breaking parameter $A_0$. However, due to the relation
$M_{1/2}^2 \geq m_0^2$, the bounds on $A_0$ and $B_0$
coming from the stability condition and the requirement
of the absence of a colour breaking minima \cite{GRZ},
and the smallness of the coefficients associated
with the $A_0$ dependence,
one gets that the correlation between $\mu$ and
$M_{1/2}$ is conserved over most of the parameter space,
\begin{equation}
\mu^2 \simeq D M_{1/2}^2 ,
\label{eq:mum121}
\end{equation}
where $D \simeq  2$.
The predictions coming from the above analysis, based
on the approximate relations
Eqs.(\ref{eq:mum12}--\ref{eq:mum121}),
must be compared with the results of the
numerical analysis, in which the running of gauge and Yukawa
couplings have been considered at the two loop level, and
all one loop threshold
corrections to the quartic couplings
and masses have been included.
The  resulting correlations between $\mu$ and $M_{1/2}$
are depicted in Fig. 3.a and Fig. 3.b, which are
in good agreement with the analytical results, although the
coefficient $D$ in Fig. 3b is somewhat smaller than the
analytical prediction, Eq.(\ref{eq:mum121}).

The information above may be used to get a further understanding
of the properties of Fig. 2. The lower bound on $M_{1/2}$,
for instance, follows from the condition $m_1^2 - m_2^2 > M_Z^2$,
which yields
\begin{equation}
M_{1/2} > \frac{M_Z}{\sqrt{\alpha}} ,
\label{eq:lowm12}
\end{equation}
where, as we said above, $\alpha \simeq 0.2$ for $M_t
\simeq 190$ GeV and $\tan\beta = 55$, while
$\alpha \simeq 0.1$ for $M_t \simeq 150$ GeV
and $\tan\beta = 38$. From Fig. 2 we
observe that, although the lower limit
on $M_{1/2}$ for $M_t = 150$ GeV is well described by
Eq.(\ref{eq:lowm12}), the one for $M_t = 190$ GeV is
somewhat higher than the predicted one from Eq.(\ref{eq:lowm12}).
This difference is  a reflection of the size of the
one loop
radiative corrections to the quartic couplings, which grow
with the fourth power  of the top quark mass and were
ignored for the obtention of Eq.(\ref{eq:lowm12}).

As we explained above, the condition $m_1^2 - m_2^2 > M_Z^2$
also excludes the points with $m_0 \geq M_{1/2}$. Furthermore,
low values of $m_0$, although consistent with the condition of
radiative breaking induce large mixings in the stau sectors
which yield stau masses lower than the neutralino ones.
The fact that low values of $m_0$ leads to a stau lighter
than the neutralinos was already noticed in Ref. \cite{AS}.
In the Figures we impose the condition of a neutral supersymmetric
particle to be the lightest one as an additional experimental
constraint. Under these conditions, the lightest supersymmetric
particle is always a bino, with mass $M_{\tilde{B}} \approx 0.4
M_{1/2}$. In order to get a quantitative understanding of
the lower limit on $m_0$, we recall that, ignoring small tau
Yukawa coupling effects,
the left and right
slepton mass parameters are given by \cite{Ibanez},
\begin{equation}
m_L^2 \simeq 0.5 M_{1/2}^2 + m_0^2, \;\;\;\;\;\;\;\;\;\;\;
m_R^2 \simeq 0.15 M_{1/2}^2 + m_0^2,
\end{equation}
while the mixing term for large $\tan\beta$ is dominated by
the $\mu$ parameter
\begin{equation}
m_{LR}^2 \simeq - h_{\tau} \mu v_2.
\end{equation}
Using the fact that, at energies of the order of $M_Z$,
$h_b/h_{\tau} \approx 1.7$, and the bottom - top unification
condition, the condition $m_{\tilde{\tau}} > M_{\tilde{B}}$
approximately yields,
\begin{equation}
m_0^2 \geq -0.15 M_{1/2}^2 +  \sqrt{ \left( 0.15 M_{1/2}^2
\right)^2 + \mu^2 m_t^2/3 }.
\label{eq:lowm0}
\end{equation}
Recalling Eqs.(\ref{eq:mum12}) and (\ref{eq:mum121}), and
using Eq.(\ref{eq:lowm0}), one can get an understanding
of the  $m_0$ region, indicated
as experimentally excluded in the Figs. 2.a and 2.b
(see also Fig. 10).

Close to the infrared quasi fixed point solution
the condition $m_3^2 \simeq 0 $ yields
$B \approx 0$ (from Eqs. (\ref{eq:mum12}),(\ref{eq:mum121})
$\mu^2>M^2_{1/2}>M^2_Z$), i.e.
\begin{equation}
\delta \equiv B_0 - \frac{ 6 A_0}{7}
\simeq - 1.4 M_{1/2}
\label{eq:dem12}
\end{equation}
In the numerical analysis, we studied the
correlations between
$A_0/M_{1/2}$ and
$B_0/M_{1/2}$, and compared it with the results coming
from Eq.(\ref{eq:dem12}). The results are depicted in Fig.4.a.
The numerical results confirm in a good degree the analytical
expectations. Analogously, for $Y/Y_f \simeq 0.6$, we obtain
\begin{equation}
\frac{B_0}{M_{1/2}} - \frac{0.5 A_0}{ M_{1/2}} = - 0.6.
\label{eq:dem121}
\end{equation}
The correlation, resulting in this case from
the numerical analysis
is depicted in Fig.4.b, being
in good agreement with Eq.(\ref{eq:dem121}), too.

The strong correlation between the parameter $\delta$ and
$M_{1/2}$, together with the $\mu$ - $M_{1/2}$ correlation,
Eqs. (\ref{eq:mum12}) - (\ref{eq:mum121}),
implies also
a strong correlation between $\mu$ and $\delta$.
The numerical correlation is presented in Figs. 5.a and 5.b,
for which we chose to plot the GUT scale parameter $\mu_0$
instead of the renormalized parameter $\mu$. From Figs. 3 and
5 we can hence obtain
also information about the  relation between $\mu$ and $\mu_0$,
which agrees well with the analytical prediction, Eq. (\ref{eq:mu}).

Observe that the condition $m_3^2 \simeq 0$  is a property
of the radiative breaking solutions with large values of
$\tan\beta$ and a not too heavy supersymmetric spectrum,
and in this sense is independent of the
condition of unification of top and bottom quark Yukawa
couplings. Since very low values of $\mu$
($\mu\approx 0$)
are not consistent
with the condition of radiative breaking of the electroweak
symmetry,  equations analogous to
Eqs.(\ref{eq:dem12}) and (\ref{eq:dem121}) will be obtained
even if we relax the bottom-top Yukawa unification condition.
We exemplify this by taking two solutions with
$Y_t(0)/Y_b(0) \simeq 2$ and large values of $\tan\beta$ and
studying the numerical solutions. The resulting
correlations are depicted in Fig. 4.c and 4.d.

When the condition of unification of bottom and top Yukawa
couplings is relaxed,
 however, large values of $M_{1/2}$ are
not longer needed to get the necessary hierarchy between
$m_1^2$ and $m_2^2$. As the bottom and tau Yukawa
couplings decrease compared with the top one, the coefficients
$\alpha$ and $\beta$, Eq.(\ref{eq:diff2}),
increase, $\beta$ becoming positive for
$Y_t(0)/Y_b(0) > 1.6$. Hence, for
$Y_t(0)/Y_b(0) > 1.6$, acceptable radiative breaking solutions may
be also obtained by taking large values of $m^2_0 \gg M^2_{1/2}$.
For these solutions
the strong correlation between $\mu$,
and $M_{1/2}$ is lost,
together with the hierarchical relation between $M_{1/2}$
and $m_0$. These results are depicted in Figs. 2.c and 2.d,
3.c and 3.d, and 5.c and 5.d,
where $Y_t(0)/Y_b(0) \simeq 2$.

In summary, in general the condition of radiative breaking
of the electroweak symmetry with large values of
$\tan\beta$ implies a strong correlation between the parameters
$M_{1/2}$ and $\delta$.
This correlation is a reflection of, in principle, a strong
degree of fine tuning,
implied by the condition $m_3^2 \approx 0$.
However, it
is tempting to speculate that this correlation has some
fundamental origin, what would imply the necessity of
redefining the naive fine tuning criteria.

If the top quark--bottom quark Yukawa coupling unification
is required, the parameters $M_{1/2}$ and $\delta$ are
also strongly
correlated with the supersymmetric mass parameter $\mu$.
These properties do not strongly depend on the
proximity to the infrared fixed point solution, although
the exact value of the coefficient relating the different
parameters and the strength of the correlation
does depend on the top quark Yukawa coupling value.
Radiative breaking of the electroweak symmetry is driven by
the gaugino mass $M_{1/2}$ and $M^2_{1/2} > m^2_0 > M^2_Z$.
For a large enough departure from the exact top quark -- bottom
quark Yukawa unification ($Y_t(0)/Y_b(0) > 1.6$)
solutions with radiative breaking driven by $m^2_0$
are also possible, for which  both the correlation
between $\mu$ and $M_{1/2}$ and the hierarchical
relation between $M_{1/2}$ and $m_0$ are destroyed.

\section{Radiative Corrections to $M_b$ and $M_{\tau}$
         and the Predictions for the Top Quark Mass}

Fig.1 summarizes the predictions for the top quark mass
as a function of $\alpha_3(M_Z)$ for given values of
$\tilde{M}_b$, which follow from unification of the three
Yukawa couplings. As explained in Section 2, the pole mass
$\tilde{M}_b$ is obtained from the unification condition
in the case in which the supersymmetric one--loop corrections
to the bottom mass are ignored (i.e.\ it includes only QCD
corrections). In this section we calculate the supersymmetric
one--loop corrections to the bottom mass in the model with
radiative breaking. For large values of
$\tan\beta$, they are not
only large but, due to the strong correlations between the
soft supersymmetry breaking parameters present in the
large $\tan\beta$ solutions with $Y_t \simeq Y_b$,
for fixed $\tan\beta$ they are
almost constant in the whole parameter space allowed by
radiative breaking. Thus, in the first approximation,
for fixed $\tan\beta$ and $M_t$, $M_b=\tilde{M}_b+\hbox{const}$.
If this value of $M_b$ is in the range of the experimentally
acceptable values for the physical bottom mass,
$M_b\simeq 4.9 \pm 0.3$ GeV, then the corresponding values of
$M_t$ and $\tan\beta$ are the predictions for the top quark
mass and $\tan\beta$, consistent with radiative breaking.
Of course, all uncertainties taken into account, the actual
prediction is a band of values for $M_t$ and $\tan\beta$.

There is a higher order ambiguity due to the choice of the
scale at which the supersymmetric one--loop corrections
are calculated. A natural choice is between the electroweak and
supersymmetric ($M_{\tilde{g}}$) scales and we choose to work
with $m_b(M_Z)$.
The corrected running bottom quark mass $m_b$ reads
\cite{Hall},\cite{Ralf}
\begin{eqnarray}
m_b = h_b v_1 \left( 1 + \Delta(m_b) \right).
\end{eqnarray}
 $\Delta(m_b)$ receives contributions coming from
bottom squark--gluino loops and top squark--chargino
loops, and is given by,
\begin{eqnarray}
\Delta(m_b) & = & \frac{2 \alpha_3}{3 \pi} M_{\tilde{g}} \;
\mu \;
\tan\beta \; I(m_{\tilde{b},1}^2, m_{\tilde{b},2}^2,
M_{\tilde{g}}^2)
\nonumber\\
& + & \frac{ Y_t }{4 \pi} A_t \; \mu \; \tan\beta  \;
I(m_{\tilde{t},1}^2, m_{\tilde{t},2}^2,\mu^2),
\label{eq:deltabm}
\end{eqnarray}
where the integral function I(a,b,c) is given by
\begin{equation}
I(a,b,c) = \frac{a b \ln(a/b) + b c \ln(b/c) + a c \ln(c/a) }
{( a - b )( b - c )( a - c )},
\end{equation}
with
$M_{\tilde{g}}$ and $m_{\tilde{b},i}$  ($m_{\tilde{t},i}$)
are the gluino
and sbottom (stop) eigenstate masses respectively. The integral
function may be parametrized as $I(a,b,c)
= K_I/a_{max}$, where $a_{max}$ is the maximum of the
three squared masses appearing in the functional integral
and the coefficient $K_I \simeq 0.5 - 0.9$ if there is
no large hierarchy between the three different masses.
Observe that
the minimum value of $K_I = 0.5$ is only obtained when
the three masses are equal. As we will discuss below,
for the typical values of
the mass parameters appearing in the radiative electroweak
symmetry breaking solutions, $K_I \simeq 0.6$ gives a
good approximation to the integral.

The tau mass corrections are, instead, dominated by the
bino exchange contribution, which is negligible for the
bottom quark case. Indeed,
\begin{equation}
m_{\tau} = h_{\tau} v_1 ( 1 + \Delta(m_{\tau}) ),
\end{equation}
with
\begin{eqnarray}
\Delta(m_{\tau}) & = & \frac{ \alpha_1}{4 \pi} \;
M_{\tilde{B}} \; \mu \;
\tan\beta \;
I(m_{\tilde{\tau},1}^2, m_{\tilde{\tau},2}^2,
M_{\tilde{B}}^2)
\end{eqnarray}
Observe that although the effect is expected to be small due
to the presence of the weak gauge coupling, it is partially
enhanced by the fact that the particles appearing in the
loop are lighter than in the bottom case. We will discuss
it in more detail below.

Due to the approximate dependence of $A_t$  on
$A_0$ and $M_{1/2}$, Eq.(\ref{eq:A0}), close to the fixed point there
is a strong correlation between $A_t$ and the gluino mass.
Indeed, for $Y/Y_f \simeq 1$,
\begin{equation}
A_t \simeq -\frac{2 M_{\tilde{g}}}{3}.
\label{eq:at1}
\end{equation}
For values of $Y/Y_f
\simeq 0.6$, $A_t$
 is shifted towards larger values in most of
the parameter space,
\begin{equation}
A_t \simeq - M_{\tilde{g}}.
\label{eq:At06}
\end{equation}
These correlations are observed in the numerical analysis.
The relations above, Eqs.(\ref{eq:at1}), (\ref{eq:At06}),
are only violated for
large values of $A_0$, close to the upper bound on this
quantity (For the numerical bounds on $A_0$ see Fig. 4).
Due to the minus signs in
Eqs.(\ref{eq:at1}) and (\ref{eq:At06}), there is an effective
cancellation between both bottom mass correction contributions.
Interestingly enough, due to the fact that $A_t$ is larger
when the Yukawa coupling
$Y$ is smaller,
the cancellation between both bottom mass
correction contributions for $Y/Y_f \simeq 0.6$
is similar to the one appearing for $Y/Y_f \simeq 1$.

 The bottom mass corrections become very
 relevant for large
values of $\tan\beta \geq 30$.
In order to reduce the bottom mass corrections, while fulfilling
the requirement $m_3^2 \simeq 0$, the authors of Ref.\cite{Hall}
imposed a Peccei Quinn symmetry $\mu \rightarrow 0$, which
is explicitly broken, its breakdown being characterized
by the (assumed) small parameter $\epsilon_{PQ} =
\frac{\mu}{m_{\tilde{q}}}$. They also required the presence
of an approximate continuous $R$ symmetry, present in the
limit $B \rightarrow 0$, $M_{\tilde{g}} \rightarrow 0$,
$A \rightarrow 0$, which breaking is characterized by
the (assumed) small parameter
\begin{equation}
\epsilon_R = \frac{B}{m_{\tilde{q}}} \simeq
\frac{A}{m_{\tilde{q}}} \simeq \frac{M_{\tilde{g}}}{
m_{\tilde{q}}},
\label{eq:symet}
\end{equation}
which would protect both $\tan\beta$ and the bottom mass
corrections.

However, the electroweak symmetry
radiative breaking solutions with universal soft
supersymmetry parameters at the grand unification scale
and exact top quark--bottom quark Yukawa coupling
unification
is inconsistent with the approximate preservation of
these symmetries.  Indeed, as we have discussed in the last
section, the only solutions satisfying these requirements
 are obtained for $M_{1/2}^2$ of the order of, or larger than
$m_0^2$. Under these conditions, the squark mass
is of the order of the gluino mass and not much larger than it,
as required by $\epsilon_R$.
In addition, as explained above, the mass parameter
$A_t$ is of the order of the gluino mass and,
hence,
the bottom mass corrections are not suppressed in the minimal
supergravity model.  Indeed,
due to the strong correlation between $\mu$ and
$M_{1/2}$, the $\mu$ parameter is strongly correlated with
the gluino mass, and an approximate expression for the
integrals $I(a,b,c)$ may be obtained.
Using these correlations, we obtain that the integrals are
well approximated by setting $K_I \simeq 0.6$,
\begin{equation}
\Delta (m_b) \simeq 1.2 \tan\beta
\frac{\mu}{M_{\tilde{g}}} \left( \frac{\alpha_3}
{3\pi} + \left(\frac{3}{2}\right)
\frac{Y_t A_t}{8 \pi M_{\tilde{g}}} \right)
\end{equation}
where the factor $3/2$ is to account for the fact of having written
a factor $M_{\tilde{g}}^2$ in the denominator, instead of
the appropiate factor $m_{\tilde{t},1}^2$ (The correlation
between the gluino and squark masses will be discussed in the
next section). The above expression
gives a good approximation to the bottom mass corrections
in most of the parameter space consistent with radiative breaking
of the electroweak symmetry and bottom--top Yukawa coupling
unification. Observe that, due to the strong correlations between
$A_t$ and $M_{\tilde{g}}$, and using the fact that the fixed point
value of the top and bottom quark Yukawa coupling is approximately
given by $Y_f \simeq 16 \alpha_3/21$, there is an effective
cancellation between the two contributions, which reduces the
gluino contribution by about a 30$\%$. In the more precise
numerical result this cancellation is of the order of $25\%$.
Taking this into account, and the fact that $M_{\tilde{g}}
\simeq \alpha_3 M_{1/2}/\alpha_{G}$, with $\alpha_G$ the unifying
value of the gauge couplings,
the relative bottom mass corrections
are given by
\begin{equation}
\Delta (m_b) \simeq 0.0045 \tan\beta \frac{\mu}{M_{1/2}}.
\label{eq:dmb}
\end{equation}

{}From Fig. 1 we observe that, in general, independent of
the bottom mass value, the condition of unification of
the three Yukawa couplings is such that the larger is
$\tan\beta$, the closer is the top quark Yukawa coupling
to its fixed point value. Values of the top quark Yukawa
coupling close to its fixed point are obtained for
$\tan\beta \approx 60$, while $Y_t/Y_f = 0.6$ is obtained
for $\tan\beta \approx 40$.
Once Eq.(\ref{eq:dmb}) is combined with
the numerical results shown in Fig. 3,
we obtain a relative bottom mass
correction of the order of 45 $\%$ for values of
$\tan\beta$ and the top quark mass consistent with the
fixed point, while for $Y_t/Y_f = 0.6$, the relative
bottom mass correction is of the order of $20 \%$.

An analogous procedure can be applied for the estimation
of the relative tau mass corrections. In this case,
the heaviest stau mass is of the order of 2.5 times the
bino mass, while the lightest stau mass is of the order
or somewhat larger than the bino mass, depending on the
relative value of $m_0$ with respect to $M_{1/2}$. Under
these conditions, over most of the allowed parameter space
the loop integral may be approximated
by a factor $K_I$ of the order of 0.85. One can check
that, under these conditions the
tau mass corrections are
not larger than $6 \%$ of the tau mass. Moreover, a
relatively large tau mass correction is always associated with a
large left - right stau mass mixing, for which the lightest
stau becomes the lightest supersymmetric particle. Once
the condition of a neutralino being the lightest
supersymmetric particle is imposed, the relative
tau mass corrections
are bounded to be lower than 4$\%$ over most of the allowed
parameter space.
Since, in addition, a relative tau mass variation affect less
the unification condition than a relative bottom mass
correction, the tau mass correction effects on
the top quark mass predictions are small.

For chosen values of $M_t$ and $\tan\beta$ we are now able
to calculate the physical bottom mass by running down the corrected
$m_b(M_Z)$ with the standard Model RGE to the scale $M_b$
and applying the appopriate QCD corrections, Eq.(\ref{eq:mb}).
The results are shown in Fig. 6, for the same representative
values of the top quark mass and $\tan\beta$ chosen in the
previous figures (which, as we discussed in section 4, for
the cases a) and b) correspond
to $\tilde{M}_b \simeq 5.4$ GeV and $\tilde{M}_b \simeq
5.85$ GeV, respectively). The two branches of
$M_b$ correspond to two signs of $\mu$, the lowest values
of $M_b$ corresponding to negative values of $\mu \times
M_{1/2}$ (or equivalently, positive values of $\mu \times
A_t$). From Fig. 6 and recalling the results presented in Fig.1,
we see that large corrections to the bottom mass, of the order of
30$\%$ may be used to reconcile $Y_t = Y_b = Y_{\tau}$ with
$M_t \leq 160$ GeV. Moreover, it is easy to see that, due to
the size of the characteristic corrections, for $\alpha_3(M_Z)
> 0.11$, there are no solutions with
$\tilde{M}_b < M_b$ and a top quark Yukawa coupling within
the range of validity of perturbation theory.

The above results may be used to set an upper bound
on the top quark mass as a function of the strong
gauge coupling value. This upper bound will correspond
to the maximum values of $\tan\beta$ and $Y_t$ consistent
with a physical bottom mass $M_b \simeq 4.6$ GeV. Larger
top quark mass values will correspond to lower values of
$\tilde{M}_b$ and larger values of $|\Delta m_b|$,
implying a physical bottom quark mass outside the
present experimental bounds on this quantity, $M_b < 4.6$
GeV. The upper bound may be hence estimated as follows:
For a given value of $\alpha_3(M_Z)$, $M_t$ and
$\tan\beta$, the relative
bottom mass corrections may be computed by using the
supersymmetric mass parameters allowed by
the condition of radiative breaking of the electroweak
symmetry. In addition, $\tilde{M}_b$ may be otained from
the correlations between $M_t$, $\tilde{M}_b$ and
$\tan\beta$ depicted in Fig. 1. We perform a scanning
over the values of $M_t$ and $\tan\beta$, looking for the maximum
value consistent with a physical bottom mass $M_b \geq
4.6$ GeV. This value of $M_t$ gives an estimate of the
upper bound on this quantity for this given value of
$\alpha_3(M_Z)$. The uncertainties associated with this
procedure will be discussed below.

For example, for $\alpha_3(M_Z) =
0.12$, the upper bound on the top quark mass is approximately
given by $M_t \leq 150$ GeV while the upper bound on the ratio
of vacuum expectation values is given by $\tan\beta \leq 39$.
These bounds are associated with a
bottom mass $\tilde{M}_b \simeq 5.6$ GeV.
{}From Eqs. (\ref{eq:deltabm}),
(\ref{eq:dmb}) it follows that the approximate
bottom mass corrections under these conditions
(corresponding to the lowest value of $\mu/M_{1/2} \simeq 1$)
are of the order of 18$\%$. Hence, as required for the
solution associated to the upper bound on the top quark mass,
the physical bottom mass will
be approximately equal to the lower experimental bound
on this quantity $M_b \simeq 4.6$ GeV.
Analogously, for $\alpha_3(M_Z) = 0.13 \;
(0.11)$ the upper bounds read $M_t \leq 170 \;
(130)$ GeV and $\tan\beta \leq 43 \;
(34)$, for which the bottom mass $\tilde{M}_b \simeq 6$ (5.3)
GeV. The lowest bottom mass corrections are of the order of
23$\%$ (13 $\%$)
(corresponding to a ratio $\mu/M_{1/2} \simeq 1.2$ $(0.85)$)
and hence the physical bottom mass is approximately equal to
4.6 GeV.

It is important to discuss the uncertainties on the estimate
of the top quark mass upper bounds presented above.
For the obtention of Fig. 1,
the supersymmetric spectrum has been taken to be degenerate
at a mass $M_Z$. However, the squark and gluino spectrum
arising from the bottom - top Yukawa unification condition
is heavy and hence, the top quark mass could be modified by
the supersymmetric particle threshold effects. These effects
have been estimated in Ref. \cite{Hall}. For the characteristic
spectrum obtained in these solutions, the resulting top quark
mass uncertainties are of the order of 5 - 10 GeV. In addition,
in the above we have ignored the possible effects of tau
mass corrections. The tau mass corrections are
correlated in sign with the bottom mass corrections and
their effects on the top quark mass predictions
may be hence estimated by a lowering of the relative
bottom mass corrections
in an amount of the order of the relative tau mass corrections.
A modification of the order of 3$\%$ of the relative bottom mass
corrections gives variations of the top quark mass prediction
of the order of 5 - 10 GeV, too. Finally, there is the already
discussed $\alpha_3$ - scale uncertainty in the evaluation of
the bottom mass corrections, which can also modify the top
quark mass predictions in a few percent.

 From the above discussion, it
follows that the estimate for the upper bound on the top quark
mass quoted above may be
away from the real bound in 10 - 20 GeV. However,
it is important to
remark that  even after the inclusion of these uncertainties
the allowed top quark mass values become much lower than
the values obtained for the case in which a negligible bottom
mass correction is assumed \cite{Hall1},\cite{CPW},\cite{Hall}.
Indeed, even after
the uncertainties are included, the upper bounds
on the physical top quark mass obtained above
are of the order of the lower
bounds for the same quantity for the case in which the bottom
mass corrections are negligible.

A lower bound on the top quark mass is also obtained.
However,
the lower bounds on the top quark mass is given by
$M_t \geq 120$ GeV for $\alpha_3(M_Z) = 0.13$, while
for values of $\alpha_3(M_Z) < 0.13$ the lower
bound is  below the present experimental limit on
the top quark mass.
In general, for $\alpha_3(M_Z) < 0.13$,
large values of the top quark mass $M_t \geq 180$ GeV will
be only possible for the case in which we relax the
condition of unification of the three Yukawa couplings.
For instance, a top quark mass $M_t \simeq 190$ GeV may
be achieved for $\tan\beta = 50$, for $Y_t/Y_b \simeq 2$,
$\alpha_3(M_Z) \simeq 0.125$ and
$\tilde{M}_b \simeq 5.2$ GeV. As we explained above,
since $m_0$ under these conditions may be much larger
than $M_{1/2}$, the approximate symmetries required in
Ref. \cite{Hall}, Eq.(\ref{eq:symet}), become now possible,
and hence the bottom mass corrections can
be small,  $M_b \simeq \tilde{M}_b$.

\section{Supersymmetric Particle Spectrum}

The properties of the sparticle spectrum are to a large
extent determined by the correlation of the mass
parameter $\mu$ and the soft supersymmetry breaking parameters
and their large values necessary to fulfill the condition of
radiative breaking of the electroweak symmetry (see section 4).
For instance, since large values of the parameters $\mu$
and $M_{1/2}$ are required, there will be little mixing in
the chargino and neutralino sectors. The lightest (heaviest)
chargino is given by a wino
(charged Higgsino) with mass equal to
$M_2 \simeq 0.8 M_{1/2}$ ($|\mu|$). The lightest neutralino will be
given by a bino of high degree of purity and mass
$M_{\tilde{B}} \simeq 0.4 M_{1/2}$. These issues have
been already discussed in Refs. \cite{SO10} - \cite{OP},
and survive in our more precise numerical correlation. We
will hence concentrate on the predictions which depend
stronger on the precise values of the top quark mass and
hence are more sensitive to the change on the top quark mass
predictions induced by the bottom mass corrections studied
in the previous section. In addition, we will present
an analysis of the constraint on the soft supersymmetry
breaking parameters coming from the present experimental
bounds on the $b \rightarrow s \gamma$ decay rate.

In Fig. 7 we give the behaviour of the CP odd Higgs mass
as a function of the lightest chargino mass. As we remarked
above, due to the large
values of $M_{1/2}$ and $\mu$ appearing in this scheme, the
lightest chargino is almost a pure wino, with mass
$m_{\chi^+} \simeq 0.8 M_{1/2}$, while the heaviest chargino
is a Higgsino with mass equal to $|\mu|$.
The CP odd Higgs mass squared is given by
$m_A^2 = \alpha M_{1/2}^2
+ \beta m_0^2 + const.$, where the constant term is negative.
Since $\alpha$ is positive and $\beta$ is negative, we
get an upper bound on $m_A^2$,
\begin{equation}
m_A^2 < \alpha M_{1/2}^2
\end{equation}
which is visible in the figures. Observe that the sensitivity
under top quark mass varations of this bound comes through
the dependence of the parameter $\alpha$ on $M_t$.
The largest values of $m_A$ are obtained for low values of
$m_0$, which, could lead to a
stau to be the lightest supersymmetric particle
(see section 4 and Ref. \cite{AS}).
{}From Figs. 7.a and 7.b, we
see that for the allowed parameter space consistent with
bottom - top Yukawa unification the CP odd Higgs
becomes light. Very low values of $m_A$ are, however,
excluded by experimental limits.
Moreover, the CP odd Higgs mass becomes
\begin{equation}
m_A \leq m_Q \sqrt{ \frac{\alpha}{5}}
\end{equation}
where the factor 5 comes from the strong correlation
between $m_Q$ and $M_{1/2}$ (see below).
For squark mass parameters $m_Q < 2$ TeV, as set in our study
and $\alpha \simeq 0.1$ as obtained for $M_t \simeq 150$ GeV
and $\tan\beta \simeq 38$,
the upper limit on the CP odd Higgs
mass,
$m_A^u \simeq 250$
GeV. Similar bounds on $m_A$ were obtained in Ref. \cite{AS}.
However, the upper bounds on $M_{1/2}$ in that work come from
an estimate of the constraints on the soft supersymmetry
breaking parameters coming from the relic density bound $\Omega
h^2 < 1$. Observe that in Ref. \cite{AS}, larger values of
the top quark mass were used, corresponding to the predictions
without bottom mass corrections.

 As it is shown in Figs. 7.c and 7.d, once the condition of
unification of top and bottom Yukawa couplings is relaxed, the
upper bound on the CP odd Higgs mass becomes weaker. The
behaviour of the CP odd Higgs mass simply reflects a change in
sign of the parameter $\beta$, which becomes positive for
these values of the top quark mass and $\tan\beta$.

In Fig. 8 we show the numerical solutions for the lightest
stop mass as a function of the gluino mass. We observe a
very strong correlation between these two quantities in
Figs. 8.a and 8.b, for which unification of top and bottom
Yukawa couplings holds. This is easily understood from the
behaviour of the squark mass parameters,
Eqs. (\ref{eq:sqm2}) and (\ref{eq:sqm}). Indeed,
we see that the $A_0$ and $m_0$ dependence of the mass parameters
$m_Q^2$ and $m_U^2$ is  weak,
becoming weaker for top quark mass values
close to the infrared fixed point solution. In addition,
values of $m_0 \geq M_{1/2}$ are forbidden by the radiative
electroweak symmetry breaking condition. Hence,
$m_U^2 \simeq m_Q^2 \simeq 5 M_{1/2}^2$ for both values of
$M_t$. The mixing is dominated by the $A_t$ term. Hence,
the lightest stop mass is given by
\begin{equation}
m_{\tilde{t}}^2 \simeq 5 M_{1/2}^2 + m_t^2
- A_t m_t.
\end{equation}
Recalling Eq. (\ref{eq:A0}), and the fact that $M_{1/2} \geq
300$ GeV, we get that in both cases $m_{\tilde{t}} \simeq
K M_{1/2}$ with $K \simeq 2.1 - 2.3$. This implies that
$m_{\tilde{t}} \simeq 0.75 M_{\tilde{g}}$, which qualitatively
describes the results shown in Figs. 8.a and 8.b.

In Figs. 8.c and 8.d we shows what happens when we depart
from the condition of exact unification. Under these conditions
large values of $m_0$ are allowed and hence the strong correlation
between the gluino mass and the lightest stop quark mass is
lost.

In Fig. 9 we plot the charged and lightest CP even Higgs
mass
spectrum. After radiative corrections, the charged Higgs
mass becomes of the order of the CP odd Higgs mass and
is hence tightly bounded from above when exact unification
of bottom and top Yukawa couplings is required. Due to
the moderate values of the top quark mass necessary to
achieve unification and the low values of the CP odd Higgs
mass, there is a large region of the allowed parameter
space where the CP even Higgs mass becomes lighter than
$M_Z$. This tendencey, however, changes  as the charged
Higgs particle mass is above 150 GeV, for which the
CP even Higgs mass reaches acquires a maximum value,
which varies only slightly for larger values of the
charged Higgs mass. This behaviour is a general
feature of the large $\tan\beta$ solutions and do not
depend on the condition of unification of bottom and
top Yukawa couplings. From Fig. 9.b, we observe that
for a top quark mass $M_t
\simeq 150$ GeV and $\tan\beta \simeq 38$,
the upper bound on the CP even
Higgs mass,  $m_h \leq 110$ GeV

In Fig. 10 we present  the lightest stau mass spectrum
as a function of the lightest chargino one. In
Figs. 10.a and 10.b it is easy to identify the
region excluded by the requirement of the lightest
stau being heavier than the bino. As we discussed
before, since for these cases
the mixing in the chargino sector is small, the
lightest chargino is  almost a pure wino with mass
$m_{\tilde{\chi}^+} \simeq 0.8 M_{1/2} \simeq
2 M_{\tilde{B}}$. Hence, this requirement implies that
the stau mass should be larger than approximately
a half of the lightest chargino mass.  Figs. 10.c
and 10.d show what happens when we depart from
the condition of exact Yukawa unification,
the larger
values of $m_{\tilde{\tau}}$ being
associated with larger
values of the soft supersymmetry breaking parameter
$m_0$.

\subsection{Experimental constraints coming from
$b \rightarrow s\; \gamma$}

In our discussion above, we have not adressed the experimental
constraints coming from the bounds on the $b \rightarrow s
\gamma$ decay rate. These bounds can be very relevant in
defining the allowed parameter space, particularly in models
with a large hierarchy between the squark and CP odd Higgs
masses \cite{BBMR}-\cite{BG},
as occur for the large $\tan\beta$ scenario when
the condition of unification of top and bottom Yukawa couplings
is required. In addition, for large values of $\tan\beta$
there is an enhancement of the
chargino - exchange contribution, with similar physical origin
as the one that enhances the bottom mass corrections for
this case.  The gluino-exchange contributions
are also enhanced, although they
are still much lower than the chargino - exchange ones \cite{Bor}.

For large values of $\tan\beta$, hence, the chargino
contributions becomes sizeable even for a characteristic
squark and Higgsino
spectrum $m_Q \simeq m_{\tilde{\chi}} \simeq
{\cal{O}}$(1 TeV), as appears in the model under study.
The sign of
these contributions, as happens with the bottom mass
corrections, depend on the sign of the product of the
parameters $\mu$
and $A_t$, related to the chargino
masses and the eigenstate stop mass splitting, respectively.

We have set a calculation of the $b \rightarrow s \gamma$
rate, following the procedure suggested in Ref. \cite{BG}.
The gluino contributions were neglected, and
the mixing of the first and second generation up squarks
was ignored.
We computed the rate numerically, according to the
results presented in Refs. \cite{BBMR} - \cite{BG}. For
the allowed soft supersymmetry breaking parameters required
for a radiative breaking of $SU(2)_L \times U(1)_Y$ with
exact unification of top and bottom Yukawa couplings,
$M_t \simeq 150$ GeV and  $\tan\beta \simeq 38$,
we find, \\
{}~\\
a) For positive values of $A_t \times \mu$, the rate is
enhanced as compared to the one of the Standard Model
with one extra Higgs doublet.\\
{}~\\
b) For negative values of $A_t \times \mu$, the rate
is smaller as the one of the Standard Model with one
extra  Higgs doublet. The chargino contributions partially
cancel the charged Higgs ones  and the rate  becomes
of the order of the Standard Model one  for squark
and Higgsino masses above a lower bound, which is
of the order of 300 GeV. \\

As we discussed above, if we insist in exact bottom -
top Yukawa unification, positive values of $A_t \times
\mu$ are needed in order to get the right values for
the bottom mass.
This means that the lower bounds on the
CP odd mass will become stronger than in the case of
the Standard Model with one additional Higgs doublet.
In order to estimate a bound on the charged Higgs mass and
the soft supersymmetry
breaking parameters, however, the uncertainties in the
computation of the rate
$b \rightarrow s \gamma$ must be
addressed.  As discussed in Ref. \cite{PB}, by far the
largest uncertainty in the $b \rightarrow s \gamma$ rate
is the one coming from the choice of the renormalization
scale of the Wilson coefficients. This uncertainty is
as large as  20 - 30 $\%$ of the computed rate \cite{PB}.
Taking this
uncertainty into account, conservative bounds on the
soft supersymmetry breaking parameters may be obtained
by demanding \cite{PB}
\begin{equation}
BR^{theor}[B \rightarrow X_s \gamma] - 2 \times  \epsilon
< BR^{u,exp},
\label{eq:BP}
\end{equation}
where $\epsilon$  is the theoretical error bar and
$BR^{u,exp} \simeq 5.4\; 10^{-4}$ is
the experimental upper bound on this
branching ratio \cite{BSGAB}.

We have computed the bounds on our model according to
Eq.(\ref{eq:BP}) and the uncertainties estimated in
Ref. \cite{PB}. For $M_t \simeq 150$ GeV, $\tan\beta
\simeq 38$, $A_t \simeq - M_{\tilde{g}}$, $\mu \simeq
- M_{1/2}$ and $m_{H^{\pm}}^2 \simeq \alpha M_{1/2}^2
+ \beta m_0^2$, as approximately required by exact
unification with $M_b \simeq 4.6$ GeV,
we find a lower bound on $M_{1/2}$ which
is approximately given by 550 GeV. If the theoretical
uncertainties are, instead, considered at the one
sigma level, the lower bound on $M_{1/2}$ is approximately
700 GeV. No significant bound on $m_0$ arises for the
model under study.

Hence, if the above described estimate of theoretical
uncertainties is correct, $b  \rightarrow
s \gamma$ will have
relevant implications for the model under study. Indeed,
if we insist in a squark spectrum below a few TeV, the
squark, chargino and Higgs spectrum will be very well
defined as follows from Figs. 7 - 9 and the lower bounds
on $M_{1/2}$ given above. A more complete study of the
rate $b \rightarrow s \gamma$, including higher loop
effects to cancel the large  uncertainties
in the rate computation will
be necessary, however, before a definite statement in this
direction can be made.

\section{Conclusions}

In this work, we have studied the conditions for a proper
breakdown of the electroweak symmetry in a minimal
supersymmetric model with unification of the three
third generation fermion Yukawa couplings and universal
soft supersymmetric parameters at the grand unification
scale. We have shown that the condition of radiative
electroweak symmetry breaking implies strong correlations
between the different soft supersymmetry breaking paramters.
These correlations have implications in the supersymmetric
particle spectrum of the theory, which strongly depends
only on the universal soft supersymmetry breaking
gaugino mass $M_{1/2}$. Moreover, a minimum value of
$M_{1/2} \geq 300$ GeV,
is implied by the condition of radiative breaking of
$SU(2)_L \times U(1)_Y$,
leading to a lower bound on the squark and
gluino masses of the order of a few hundred GeV.
The lightest supersymmetric particle becomes hence
a bino
with mass $M_{\tilde{B}} \simeq 0.4 \; M_{1/2}$.
In addition, the CP odd Higgs and charged
Higgs masses become much lower than the characteristic
squark spectrum of the theory. For squark masses lower
than a few TeV, the heavy Higgs acquire masses of the order
of the weak scale.

We have shown that, due to large bottom mass corrections
induced through sparticle exchange loops, the predicted
values of the top quark mass and $\tan\beta$ are much
lower than the values previously estimated in the literature
under the assumption that these corrections were negligible.
Indeed, we have shown that, for $\alpha_3(M_Z) \simeq 0.12$,
a top quark mass
$M_t \geq 170$ GeV is difficult to achieve
within this scheme. This tight bound on the top quark
mass may only be avoided by a relaxation of either
the exact unification of the Yukawa couplings or of
the universality of the soft supersymmetry breaking
parameters at the grand unification scale.

Finally, we have shown that the large negative
bottom mass corrections
are also associated with an enhancement of the rate
of the decay
$ b \rightarrow s \gamma$ with respect to the one
of the standard model with one extra Higgs doublet.
A lower bound on the universal gaugino mass may be
hence obtained from requiring the total decay rate
to be below the experimental upper bound on this
quantity $M_{1/2} \geq 550$ GeV.
This bound depends, however, on an estimate of the
QCD uncertainties associated with the rate computation.
Consistency between this bound and the requirement of
a not too heavy supersymmetric spectrum makes the
model very predictable and, hence, easy to test experimentally.
{}~\\
{}~\\
Acknowledgements.  S.P. would like to thank P. Chankowski,
L. Hall, J. Louis, H. P. Nilles and S. Raby for useful
discussions. M.C. and C.W. would like to thank G. Giudice,
R. Hempfling and F. Zwirner
for useful discussions. M.O. and S.P. are partially
supported by the Polish Committee for Scientific Research.
The work of M.C. and C.W. is  partially
supported by the Worldlab.

\newpage
{}~\\
FIGURE CAPTIONS
{}~\\
{}~\\
Fig. 1. Top quark mass predictions as a function of the
strong gauge coupling, within the framework  of
exact unification
of the three Yukawa couplings of the third generation.
The solid lines represent  constant values of the
bottom mass $\tilde{M}_b$, equal to A) 4.6 GeV, B) 4.9 GeV,
C) 5.2 GeV, D) 5.5 GeV and E) 5.8 GeV. The dashed lines
represent constant values of  $\tan\beta$,
equal to a) 40, b) 45, c) 50, d) 55 and e) 60.
The region to the right of the dotted curve is that one
consistent with the unification of  gauge couplings and the
experimental correlation between $M_t$ and $\sin^2\theta_W(M_Z)$.
The upper dashed line represents the infrared quasi
fixed point values for
the top quark mass, for which $Y_t(0) \simeq 1$. \\
{}~\\
Fig. 2. Values of the universal gaugino mass
$M_{1/2}$ and the soft supersymmetry breaking parameter
$m_0$ consistent with
the condition of radiative electroweak symmetry breaking.
The plots are done for two characteristic values of the top
quark mass and $\tan\beta$ consistent with exact Yukawa coupling
unification: a) $M_t = 190$ GeV , $\tan\beta = 55$ and
b) $M_t = 150$ GeV, $\tan\beta = 38$, and two values for which
$Y_t(0) \simeq
2 Y_b(0)$: c) $M_t = 190$ GeV, $\tan\beta = 50$ and
d) $M_t = 150$ GeV, $\tan\beta = 30$.
The points allowed by the radiative breaking
condition, but excluded experimentally, are represented
by dots.   \\
{}~\\
Fig. 3. The same as Fig.2, but  for the  supersymmetric
mass parameter $\mu$ and the universal gaugino mass $M_{1/2}$.\\
{}~\\
Fig. 4. The same as Fig. 2, but for the ratios of soft breaking
parameters $A_0/M_{1/2}$ and $B_0/M_{1/2}$.\\
{}~\\
Fig. 5. The same as Fig. 2, but for the supersymmetric mass
parameter at the grand unification scale $\mu_0$ and the
soft supersymmetry breaking parameter $\delta$.\\
{}~\\
Fig. 6. Predictions for the physical bottom mass as a
function of the gluino mass for the same characteristic
values of $M_t$ and $\tan \beta$
as in Fig. 2. \\
{}~\\
Fig. 7. The same as Fig. 2, but for the CP - odd Higgs
and the lightest chargino masses. \\
{}~\\
Fig. 8. The same as in Fig. 2, but for the lightest stop
and the gluino masses. \\
{}~\\
Fig. 9. The same as Fig. 2, but for the charged
and the lightest CP - even Higgs masses. \\
{}~\\
Fig. 10. The same as Fig. 2 but for the lightest stau
and the lightest chargino masses.

\end{document}